\documentclass[usenatbib]{mn2e}
\usepackage{natbibmnfix,graphicx,times}

\newcommand{\kel}{\mbox{ K}}

\newcommand{\Mpc}{\mbox{ Mpc}}

\newcommand{\secinv}{\mbox{ s$^{-1}$}}
\newcommand{\hunits}{\mbox{ km s$^{-1}$ Mpc$^{-1}$}}
\newcommand{\bq}{\begin{equation}}
\newcommand{\eq}{\end{equation}}
\newcommand{\bqa}{\begin{eqnarray}}
\newcommand{\eqa}{\end{eqnarray}}

\newcommand{\lya}{Ly$\alpha$ }

\def\VEV#1{\left\langle #1\right\rangle} 
\newcommand{\bxion}{\bar{x}_i}

\title[The ionizing background during overlap]{The ionizing background at the end of reionization}

\author[Furlanetto \& Mesinger]{Steven R.  Furlanetto$^1$\thanks{sfurlane@astro.ucla.edu} \& Andrei Mesinger$^2$ \\
$^1$Department of Physics and Astronomy, University of California, Los Angeles, CA 90095, USA \\
$^2$Hubble Fellow; Department of Astrophysical Sciences, Princeton University, Princeton, NJ 08544, USA}
\begin{document}

\maketitle

\begin{abstract}
One of the most sought-after signatures of reionization is a rapid increase in the ionizing background (usually measured through the \lya optical depth toward distant quasars).  Conventional wisdom associates this with the ``overlap" phase when ionized bubbles merge, allowing each source to affect a much larger volume.  We argue that this picture fails to describe the transition to the post-overlap Universe, where Lyman-limit systems absorb ionizing photons over moderate lengthscales ($\la 20$--$100 \Mpc$).  Using an analytic model, we compute the probability distribution of the amplitude of the ionizing background throughout reionization, including both discrete ionized bubbles and Lyman-limit systems (parameterized by an attenuation length).  We show that overlap does \emph{not} by itself cause a rapid increase in the ionizing background or a rapid decrease in the mean \lya transmission toward distant quasars.  More detailed semi-numeric models support these conclusions.  We argue that rapid changes should instead be interpreted as evolution in the attenuation length itself, which may or may not be directly related to overlap.
\end{abstract}
  
\begin{keywords}
cosmology: theory -- intergalactic medium -- diffuse radiation
\end{keywords}

\section{Introduction} \label{intro}

In the last several years, the cosmological community has made an enormous effort to measure the reionization history of the intergalactic medium (IGM).  But the picture remains murky: some evidence (principally the cosmic microwave background, or CMB) points toward a mostly-ionized universe at $z \ga 9$ \citep{page07, komatsu08, dunkley08}, but other observations (principally from quasar absorption) are usually taken to imply that reionization ends only at $z \approx 6$ \citep{fan02, mesinger04, fan06, mesinger07-prox}.  Many other techniques cannot yet distinguish between early and late reionization scenarios (e.g., \citealt{totani06, kashikawa06, ota07, mcquinn07-lya, mcquinn08-damp}; see also \citealt{fan06-review} for a review).

While there is not necessarily a contradiction between the CMB and quasar measurements -- reionization may simply be relatively extended, which would hardly be surprising from a theoretical standpoint (e.g., \citealt{cen03-letter, wyithe03-letter, furl05-double, iliev07-selfreg}) -- the tension does call for a critical examination of the existing evidence.  The most widely-recognized point in favor of late reionization is the apparent rapid decrease in the mean \lya transmission toward quasars at $z \ga 6$ \citep{fan01, fan02, white03, fan06}.  There are two aspects to this.  First, a complete \citet{gunn65} absorption trough appears toward some quasars, although the enormous optical depth of that line means that this still only requires a small neutral fraction.  Second, there appears to be a substantial steepening of the amount of absorption beyond $z \sim 6$ \citep{fan06}.  However, the latter conclusion has been challenged on empirical grounds \citep{songaila02, songaila04, oh05, becker07}; unfortunately, known lines of sight are sparse enough that no clear resolution has emerged (e.g., \citealt{lidz06}).  

Here, we take a complementary approach and examine the theoretical underpinnings of the conclusion that such a rapid change implies a detection of ``reionization."  The crux of this argument is that the moment of ``overlap" (i.e., the point at which ionized bubbles merge into much larger units, usually considered to be the ``end" of reionization) must be accompanied by a sudden, rapid increase in the amplitude $\Gamma$ of the ionization rate.  Qualitatively, this expectation comes from percolation models of the reionization process:  when ionized bubbles (presumed to be transparent to ionizing photons) overlap, many more sources illuminate any given patch, so $\Gamma$ should increase rapidly.  Quantitatively, the first self-consistent simulation of reionization showed precisely such a jump \citep{gnedin00}.  In that simulation, $\Gamma$ evolved much more sluggishly both before overlap (when ionized regions grew slowly because the photons had to ionize fresh material) and afterwards (when the ionizing photons were already able to propagate large distances), so overlap appeared to be clearly defined.

However, this picture does not match properly onto the well-understood post-reionization Universe, where dense ``Lyman-limit systems" (LLSs) absorb ionizing photons over relatively small distances (e.g., $\sim 110 \Mpc$ at $z=4$; \citealt{storrie94, miralda03}, probably falling to $\la 30 \Mpc$ by $z=6$; \citealt{lidz07}).  Once ionized regions grew beyond the mean separation of these systems, LLSs (rather than the edges of ionized bubbles) controlled the mean free path of ionizing photons \citep{furl05-rec, gnedin06} and by extension the effective horizon to which sources could be seen.  Because reionization is so inhomogeneous, many ionized regions will reach this size ($\ga 20 \Mpc$) well before complete overlap \citep{barkana04, furl04-bub}; there must in fact be a gradual transition from the ``bubble-dominated" ionization topology characteristic of reionization to the ``web-dominated" topology characteristic of the post-reionization Universe \citep{furl05-rec}.  

Here we ask whether overlap \emph{must} be accompanied by a rapid increase in $\Gamma$ and, conversely, whether such an increase is a robust signal of overlap.  We also examine the implications of recent reionization models for the \lya forest observations conventionally used to argue for late reionization.  Unfortunately, numerical simulations do not yet have the dynamic range to sample the large ($\sim 100 \Mpc$) scales necessary for reionization and simultaneously predict detailed properties of the \lya forest (though see \citealt{gnedin06} for a detailed study of LLSs and the \lya forest during reionization in a $8 h^{-1} \Mpc$ box).  Thus we will use a simple analytic model that incorporates both the discrete ionized bubbles and intervening absorption by LLSs to show that, during the end stages of reionization, $\Gamma$ is primarily controlled by the mean separation of LLSs, which may not evolve rapidly at the moment of overlap -- and, more importantly, may continue evolving long afterwards.

In our numerical calculations, we assume a cosmology with $\Omega_m=0.26$, $\Omega_\Lambda=0.74$, $\Omega_b=0.044$, $H_0=100 h \hunits$ (with $h=0.74$), $n=0.95$, and $\sigma_8=0.8$, consistent with the most recent measurements \citep{dunkley08,komatsu08}.   Unless otherwise specified, we use comoving units for all distances.

\section{Method}
\label{method}

We wish to compute the probability distribution $f(J)$ of the angle-averaged specific intensity of the radiation background at the HI ionization edge, $J$.  Its value at a given point depends on four basic parameters:  the number density of ionizing sources, $n_i$; the characteristic luminosity of each source, $L_\star$; the size of the local ionized bubble, $R$; and the attenuation length for ionizing photons within each bubble, $r_0$.  This last quantity is determined by the spacing of LLSs in the mostly ionized IGM \citep{miralda03}.

We first suppose that $R$ is prescribed and compute $f(J)$ within that single ionized region.  For the sources, we assume that every dark matter halo in the region with a virial temperature above the minimum level for atomic cooling ($T_{\rm vir} > 10^4 \kel$; \citealt{barkana01} and references therein) hosts a single galaxy with luminosity proportional to its mass, although the mass function is steep enough that the luminosity-mass relation makes little difference to our results.  Our qualitative results are also unaffected if only a fraction of galaxies actively form stars (and hence do not produce ionizing photons) at any given time; as we will see below, random fluctuations are small, even if most galaxies are quiescent.

For our purposes, we can divide the ionized bubbles into two limiting regimes.  Small bubbles (usually appearing early in reionization) have $R \ll r_0$.  In this case, we can let $r_0 \rightarrow \infty$ (or, equivalently, the optical depth $\tau=r/r_0$ experienced by a photon within that bubble vanishes).  We then compute $f_{\tau=0}(J)$ following \citet{zuo92a}, with a simple extension to account for the luminosity function of the sources (as in \citealt{meiksin04}).  Assuming that the total number of sources in the bubble is Poisson-distributed with mean $\bar{N} = (4\pi/3) n_i R^3$ (a reasonable approximation according to N-body simulations; \citealt{casas02}),
\bqa
f_{\tau=0}(j) & = & {1 \over \pi} \int_0^{\infty} ds \, \exp \left[ \bar{N} \int dx \, \phi(x) \kappa_1(sx) \right] \nonumber \\
& & \times \cos  \left[ -s j + \bar{N} \int dx \, \phi(x) \kappa_2(sx) \right] .
\label{eq:jdist_R}
\eqa
Here $j=J/J_\star^{\tau=0}$, $J_\star^{\tau=0}=L_\star/(4 \pi R)^2$, $L_\star$ is the mean luminosity of the sources, $x=L/L_\star$, $\phi(x)$ is the source luminosity function normalized so that $\int dx \phi(x)=1$, 
\bqa
\kappa_1(t) & = & \cos(t) - 1 - 2 t \, {\rm Im} \, g(t) \\
\kappa_2(t) & = & \sin(t) + 2 t \, {\rm Re} \, g(t)
\eqa
and 
\bq
g(t) = \int_0^1 du \, e^{it/u^2}.
\label{eq:gdefn}
\eq
In this limit, the mean background is $\VEV{j}_{\tau=0}=3 \bar{N}$, or $\VEV{J} \propto \bar{N} J_\star^{\tau=0} \propto L_\star R$.

In the later stages, most bubbles become much larger than the attenuation length set by embedded LLSs ($r_0 \ll R$), which then absorb most of the ionizing photons \citep{furl05-rec}.  In this case, we can approximate the ionizing background by taking the limit $R \rightarrow \infty$ \citep{meiksin04}, which gives
\bqa
f_{R=\infty}(j') & = &  {1 \over \pi} \int_0^{\infty} ds \, \exp \left[ -s \bar{N}_0 \int dx \, x \phi(x) \, {\rm Im} \, G(sx) \right] 
\nonumber \\
& & \times \cos \left[ -sj' + s \bar{N}_0 \int dx \, x \phi(x) {\rm Re} \, G(sx) \right],
\label{eq:jdist_r0}
\eqa
where $\bar{N}_0 = (4 \pi/3) n_i r_0^3$, $j'=J/J_\star^{R=\infty}$, $J_\star^{R=\infty} = L_\star/(4 \pi r_0)^2$,
\bq
G(t) = \int_0^\infty du \, \tau^3(u) e^{itu},
\label{eq:Gdefn}
\eq
and $u=e^{-\tau}/\tau^2$.  In this limit, the mean background is $\VEV{j'}_{R=\infty} = 3 \bar{N}_0$, or $\VEV{J} \propto \bar{N}_0 J_\star^{R=\infty} \propto L_\star r_0$.

\citet{meiksin04} describe how to obtain the full distribution for arbitrary $R$ and $r_0$; however, we find that the two limiting cases above are reasonable approximations to the true distribution (and much simpler to compute), provided that they are renormalized to have the proper mean value that includes \emph{both} attenuation and the finite bubble size \citep{meiksin04}:
\bq
\VEV{j'} = 3 \bar{N}_0 (1 - e^{-R/r_0}).
\label{eq:jmean}
\eq
The approximation is effective because reasonable attenuation lengths at the end of reionization ($R \ga 10 \Mpc$) contain huge numbers of sources, for which the limiting forms converge to each other.  For a given bubble, we therefore use equation~(\ref{eq:jdist_R}) if $R<r_0$ or equation~(\ref{eq:jdist_r0}) otherwise, rescaled by a constant factor so that $f(j)$ has the proper mean.

\begin{figure}
\begin{center}
\resizebox{8cm}{!}{\includegraphics{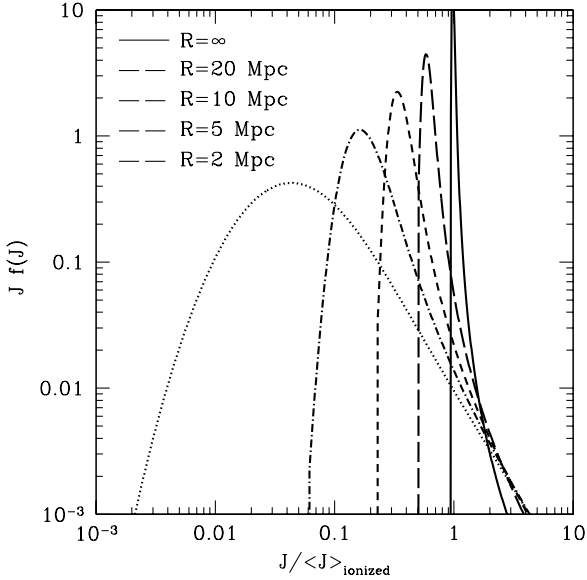}}\\%
\end{center}
\caption{Distribution of $J$ relative to its mean value in a fully-ionized IGM, $\VEV{J}_{\rm ionized}$.  All curves assume $z=6.5$ and $r_0=20 \Mpc$.  The solid curve is for a fully-ionized universe; the others assume discrete bubbles with $R=2,\,5,\,10,$ and $20 \Mpc$, from left to right, or $\bar{N}=(0.155,\,0.241,\,1.93,\,15.4) \times 10^4$.}
\label{fig:bub_ex}
\end{figure}

Figure~\ref{fig:bub_ex} shows some example distributions for bubbles at $z=6.5$ (with $n_i$ computed from the \citealt{press74} distribution).  Note that we have normalized $J$ to its mean value in a fully ionized IGM  (taking $R \rightarrow \infty$). Clearly the distributions narrow rapidly as more sources become visible (i.e., as $R$ increases); this is because the fractional variation in the source counts goes like $1/\sqrt{\bar{N}} \sim 1/R^{3/2}$ (ignoring absorption).  However, the peak (and the mean) evolve much less rapidly.  When $R \ll r_0$, $\VEV{J} \propto R$, but the evolution slows once $R \sim r_0$.  Indeed, for a constant $r_0$, the mean amplitude increases by only a factor $\sim 1.6$ from $R=r_0$ to $R=\infty$, as the additional sources are already significantly attenuated.

There are three shortcomings of this simple model.  First, our rescaling via equation~(\ref{eq:jmean}) does not exactly preserve the shape of the $J \gg \VEV{J}$ tail, which corresponds to points very near to a single source; in this limit $f(J) \propto J^{-5/2}$ and all the curves should converge to the solid one.  Fortunately, precisely because this tail describes the immediate neighborhood of sources, it is not important for our argument:  we are only concerned with the ionizing background in the diffuse IGM, where transmission in the \lya forest appears.  In practice, at the ionized fractions of interest most of the volume is filled by large bubbles ($R>r_0$) for which the rescaling is quite accurate.

More seriously, these curves assume that the galaxy density is Poisson-distributed around the \emph{universal} mean value; in reality, galaxy clustering causes variations in the number counts above and beyond the Poissonian fluctuations in the halo number counts that we include (c.f. \citealt{wyithe06-var, mesinger08-ib}).  These have fractional amplitude $\sim \bar{b}_{\rm i} \sigma(r_0)$ in the fully-ionized limit, where $\bar{b}_{\rm i}$ is the mean bias of ionizing sources and $\sigma^2(r_0)$ is the variance in the dark matter field smoothed on a scale $r_0$; for $z=6$ and $r_0=20 \Mpc$, we have $\bar{b}_{\rm i} \sigma(r_0) \sim 0.25$.  To obtain $f(J)$ throughout the universe, the solid curve in Fig.~\ref{fig:bub_ex} must be convolved with the underlying distribution of $\bar{N}$ determined by clustering, which broadens it substantially \citep{mesinger08-ib}.  Fortunately, this only strengthens our conclusions (see the discussion in \S \ref{obs}), so we defer a detailed model of it to future work.

Finally, we also assume a spatially constant $r_0$ throughout the Universe.  This is certainly an oversimplification, because $r_0$ must itself depend on $J$.  However, the largest variations in $J$ are across bubbles with $R \ll r_0$, for which the attenuation length does not strongly affect our calculations.  When $r_0$ is important, the variations in $J$ are relatively modest anyway, so our assumption seems a reasonable one.

\section{The Ionizing Background During Reionization}
\label{ionbkgd}

To obtain $f(J)$ during reionization, we must convolve the distributions for single bubbles in Figure~\ref{fig:bub_ex} with the distribution of ionized bubble sizes, $n_b(R)$.  We use the analytic formulation of \citet{furl04-bub} to compute the latter; note that it is determined primarily by the ionized fraction and has only a weak dependence on redshift \citep{furl05-charsize, mcquinn07, mesinger08-lya}, so our results are fairly generic.  We also use this formalism to compute the mass function of dark matter halos inside each bubble \citep{furl04-lya}; note that this procedure \emph{does} properly capture clustering on scales $R<r_0$, because it is precisely this galaxy clustering which determines $n_b(R)$. In other words, overdense regions with more galaxies are already part of larger ionized bubbles, with a correspondingly more galaxies \citep{furl04-lya}.

\begin{figure}
\begin{center}
\resizebox{8cm}{!}{\includegraphics{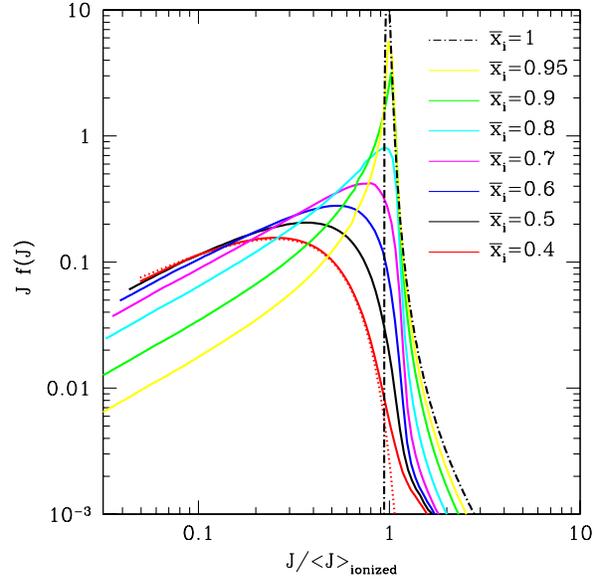}}\\%
\end{center}
\caption{Distribution of $J$ during reionization relative to its mean value in a fully-ionized IGM, $\VEV{J}_{\rm ionized}$.  The solid curves take $\bxion=0.4,\,0.5,\,0.6,\,0.7,\,0.8,\,0.9,$ and $0.95$, from left to right.  The dot-dashed curve is for a fully-ionized universe.  The dotted curve shows $f(J)$ when $\bxion=0.4$ if we ignore fluctuations inside the discrete bubbles.  All assume $r_0=20 \Mpc$ and $z=6.5$.}
\label{fig:jdistbn}
\end{figure}

Figure~\ref{fig:jdistbn} shows $f(J)$ at $z=6.5$ for a variety of ionized fractions [normalized so that $\int dJ \, f(J) = \bxion$, the mean ionized fraction].  In all cases we assume that $r_0=20 \Mpc$, consistent with the extrapolations of \citet{lidz07} from models of the \lya forest at lower redshifts.  The dot-dashed curve shows $f(J)$ once overlap is complete.  Again, the most important result is the relatively slow evolution of $f(J)$ throughout the last half of reionization; $\VEV{J}$ increases primarily because the tail toward small $J$ shrinks as more and more of the Universe is incorporated into large bubbles.  

Because the ionized bubbles reach such large sizes early in reionization (typically $\sim 4.5 \Mpc$ when $\bxion=0.5$), the radiation background incident on a typical ionized patch increases by only a factor $\sim 5$ throughout the last half of reionization:  although the size of each ionized region continues to increase until $R \rightarrow \infty$, the fixed attenuation length prevents a corresponding rapid increase in $\VEV{J}$.  Moreover, regardless of $\bxion$, some regions have $J \approx \VEV{J}_{\rm ionized}$ even though they are not particularly near any individual source.  This broad range occurs because $n_b(R)$ always contains some large bubbles, even early in reionization, and is not due to random Poisson fluctuations in the galaxy counts inside each bubble.  To see this, the dotted curve shows what happens if we ignore fluctuations \emph{within} each bubble at $\bxion=0.4$ (i.e., we set $J$ equal to its mean value for that $R$); it traces the full curve closely except in the high-$J$ tail that is due to proximity to a single source.\footnote{Imposing a duty cycle $f_d$ on star formation (and emission of UV photons), or suppressing star formation in small halos, also has no substantial effect.  Even if $f_d \sim 0.01$, significantly smaller than one would expect from the dynamical times of galaxies, the shape of $f(J)$ changes only slightly, except in the high-$J$ tail (which increases in amplitude, because each galaxy dominates a larger region around itself).}

\begin{figure}
\begin{center}
\resizebox{8cm}{!}{\includegraphics{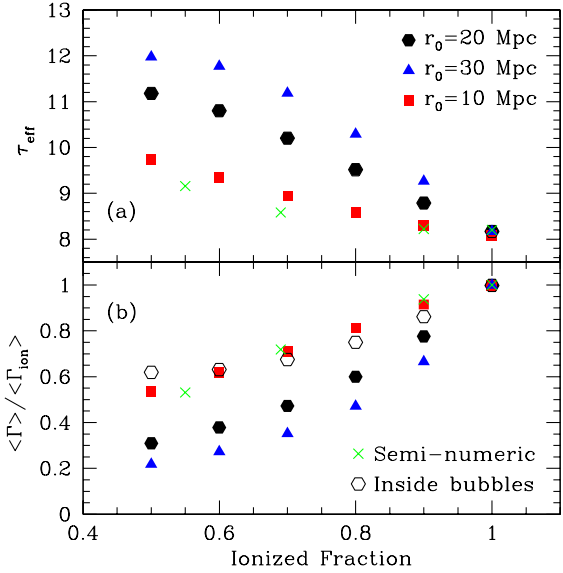}}\\%
\end{center}
\caption{Effective optical depth (\emph{a}) and mean ionizing background (\emph{b}) as a function of $\bxion$ at a constant redshift.  The filled squares, hexagons, and triangles assume $r_0=10,\,20,$ and $30 \Mpc$, respectively, at $z=6.5$.  The open hexagons show the mean ionizing background inside ionized regions, $\VEV{\Gamma_{-12}}/\bxion$, for $r_0=20 \Mpc$ at $z=6.5$.  The crosses use $f(J)$ taken from the semi-numeric simulations of \citet{mesinger08-ib} (see text). }
\label{fig:tau-z65}
\end{figure}

\section{Observable Implications}
\label{obs}

We now examine how the slow evolution of $f(J)$ affects the \lya forest.  The filled hexagons in Figure~\ref{fig:tau-z65}\emph{b} show the mean ionization rate $\VEV{\Gamma}$ for the models in Figure~\ref{fig:jdistbn};\footnote{Here we assume $\Gamma \propto J$, neglecting the variations in the mean free path for higher-energy photons because they do not contribute strongly to the total ionization rate.} the filled squares and triangles show the same sequence of ionized fractions, but with $r_0=10$ and $30 \Mpc$, respectively (approximately spanning the range of values expected from extrapolating \lya forest measurements at lower redshifts; \citealt{lidz07}).  Note that these values are averaged across the \emph{entire} Universe; the mean values \emph{within} ionized regions are shown for the $r_0=20 \Mpc$ model by the open hexagons.  Again, it is obvious that $\VEV{\Gamma}$ evolves only slowly during reionization.  The rate of evolution decreases with $r_0$, because bubbles reach the ``saturation radius" $R=r_0$ quicker \citep{furl05-rec}; beyond that point, $\VEV{J}$ can only increase by a factor $(1 - 1/e)^{-1}$ so long as $r_0$ remains fixed.

The most easily observed property of quasar \lya forest spectra is the mean transmission $\mathcal{T} \equiv \exp(-\tau_{\rm eff})$; we show the effective optical depth in Figure~\ref{fig:tau-z65}\emph{a} for the same set of models.  To compute $\tau_{\rm eff}$, we convolve the (volume-averaged) IGM density distribution of \citet{miralda00} with $f(J)$.  We assume that the local density is uncorrelated with $J$; while this is certainly not correct in detail, it is a reasonable first step because $\Gamma$ is typically dominated by distant galaxies (Olber's paradox).  Note that the absolute value of $\tau_{\rm eff}$ is somewhat sensitive to the high-$J$ tail; we truncate all of our calculations at $10 \VEV{J}_{R=\infty}$ (this essentially prevents the low-density voids that provide transmission from being extremely close to galaxies).  For concreteness, we assume that $\VEV{\Gamma_{-12}} \equiv \VEV{\Gamma}/(10^{-12} \secinv) = 0.05$ when $\bxion=1$, consistent with limits from \lya forest measurements in quasar spectra at $z \sim 6$ \citep{fan02, fan06, bolton07}.  This yields a reasonable $\tau_{\rm eff} \approx 8$ at $z=6.5$ (c.f. \citealt{fan01, fan06}).  

\begin{figure}
\begin{center}
\resizebox{8cm}{!}{\includegraphics{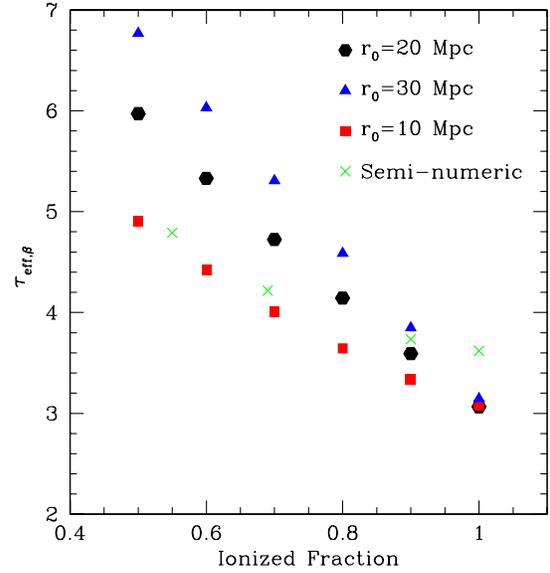}}\\%
\end{center}
\caption{Same as Fig.~\ref{fig:tau-z65}\emph{a}, except for $\tau_{{\rm eff}, \beta}$.}
\label{fig:tau_beta}
\end{figure}

Figure~\ref{fig:tau-z65} shows that there need not be a discontinuity in $\tau_{\rm eff}$ near the end of reionization:  in all of our models, the transmission evolves smoothly and (relatively) slowly as a function of ionized fraction, especially when $r_0$ is small (so that bubbles enter the saturated regime earlier).  However, the observational relevance of these results is questionable, because such large $\tau_{\rm eff}$ are probably unobservable.  Fortunately, the higher Lyman-series transitions are easier to observe; for example, Ly$\beta$ has an optical depth 6.24 times smaller than Ly$\alpha$ (although the ratio $\tau_{\rm eff,\beta}/\tau_{\rm eff}$ is closer to unity because of the convolution with the density field; \citealt{songaila02, songaila04, oh05}).  Figure~\ref{fig:tau_beta} shows $\tau_{\rm eff,\beta}$ for the same models as Figure~\ref{fig:tau-z65}\emph{a}.  This shows the same qualitative trends as for Ly$\alpha$, although the evolution is somewhat steeper, and again illustrates that no strong break is necessary at overlap.

As described above, the most important shortcoming of our model is that it ignores galaxy clustering when $R>r_0$; clustering substantially broadens $f(J)$ (but does not affect $\VEV{J}$ much).  To check the importance of this effect, we have drawn $f(J)$ from the $z=7$ semi-numeric simulations of \citet{mesinger08-ib} (in turn based on the reionization model of \citealt{mesinger07}).  These simulations include attenuation (with $r_0=20 \Mpc$), clustering, and more accurate bubble sizes.  In principle, we can also use them to compute correlations between the flux field and the underlying density field; however, the finite box size ($100 \Mpc$) means that rare voids (which account for the bulk of the transmission) may be missed and the quasi-linear treatment may not capture the full density distribution.  We therefore draw $f(J)$ from the semi-numeric simulations and convolve it with the \citet{miralda00} density distribution to compute $\tau_{\rm eff}$, shown by the crosses in Figures~\ref{fig:tau-z65} and~\ref{fig:tau_beta}.\footnote{We actually take the $z=7$ $f(J)$ from the simulations and apply them to the $z=6.5$ density field.  We then choose $\VEV{\Gamma_{-12}}=0.032$ at $\bxion=1$ so that $\tau_{\rm eff}$ is identical to that in the analytic model.  Note that, by widening $f(J)$, clustering \emph{decreases} the required mean ionizing background for a given mean transmission level.  Interestingly, it also changes the relative ratio of \lya and Ly$\beta$ transmission.  We have verified that the slow evolution persists even if we use the semi-numeric density field.}  These show \emph{even slower} evolution throughout reionization, partly because the simulated ionized regions are somewhat larger than the analytic model predicts and partly due to the inclusion of clustering when $R > r_0$.  Thus the semi-numeric approach confirms our conclusion that overlap need not be accompanied by a rapid increase in $\VEV{\Gamma}$, unless $r_0$ also increases.

\begin{figure}
\begin{center}
\resizebox{8cm}{!}{\includegraphics{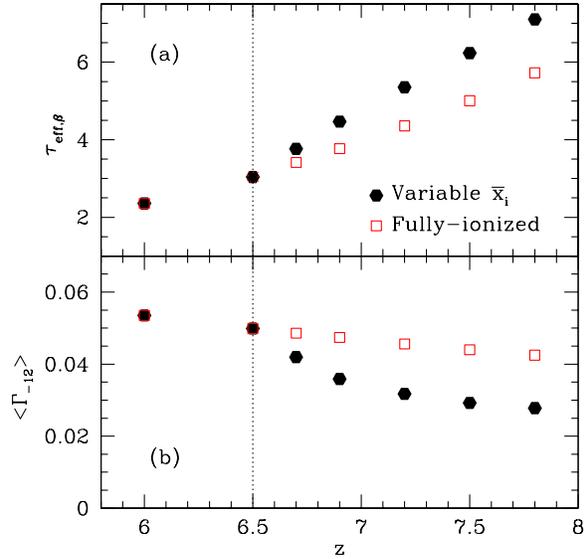}}\\%
\end{center}
\caption{Evolution of the effective optical depth in the Ly$\beta$ transition (\emph{a}) and the mean ionizing background (\emph{b}) for a scenario in which reionization ends at $z \le 6.5$ (marked by a vertical dotted line) and $r_0 \propto (1+z)^{-3}$ (filled hexagons).  Above $z=6.5$, adjacent hexagons have $\Delta \bxion=0.1$ between them (with $\bxion=0.5$ at $z=7.8$).  The open squares show an alternate model with a constant comoving emissivity and $\bxion=1$ throughout this range.}
\label{fig:tau_evol}
\end{figure}

To this point, we have fixed the mean free path and redshift to isolate how $\bxion$ affects $f(J)$.  With the filled hexagons in Figure~\ref{fig:tau_evol}, we specialize to a particular reionization scenario where $\bxion$ is proportional to the fraction of gas in galaxies, and where we set $\bxion=1$ at $z \le 6.5$.  We further assume that $r_0 = 15 [(1+z)/7.5]^{-3} \Mpc$, consistent with extrapolations from the lower redshift forest \citep{lidz07}.  Again, we normalize so that $\VEV{\Gamma_{-12}} = 0.05$ at $z=6.5$.  Even with this evolution in $r_0$, the mean transmission evolves smoothly across this entire range, and there is only a modest break in $\VEV{\Gamma}$ at $z=6.5$.

In contrast, the open squares assume $\bxion=1$ at all redshifts, with an identical $r_0(z)$ and a constant comoving emissivity.  The latter does give more transmission at all redshifts, but both models evolve quite smoothly -- there is only a slight break in the reionization model's slope at overlap compared to the fully-ionized one.  Thus overlap does not, in general, cause an obvious feature in the transmission, even for the higher-series transitions.  In fact, slow evolution is guaranteed to occur unless the mean free path itself evolves more rapidly during reionization.

\section{Discussion}
\label{disc}

We have shown that, because attenuation due to LLSs must become important before overlap, the late stages of reionization need not be accompanied by a rapid increase in $\VEV{\Gamma}$ (or $\tau_{\rm eff}$).  Thus, the conventional wisdom that a sharp increase in the ionizing background is a robust indicator of overlap must be modified:  namely, such a feature indicates only that the attenuation length $r_0$ is evolving rapidly. 

Of course, it may be that $r_0$ does evolve quickly at the end of reionization (and some theoretical calculations suggest that this does occur; \citealt{wyithe08-prox, choudhury08}).  For example, $\VEV{\Gamma}$ must (at least to some degree) control the ionization structure of LLSs and hence $r_0$ \citep{miralda05, schaye06}.  As $\VEV{\Gamma}$ increases, LLSs will shrink, increasing $r_0$ and hence $\VEV{\Gamma}$, creating a positive feedback loop.  Thus even a slow initial increase in the emissivity and/or mean free path at the end of reionization may spiral into a relatively rapid change in $\VEV{\Gamma}$.  Unfortunately, our understanding of LLSs is not yet advanced enough to determine the effectiveness of such a feedback loop; for now, we can only say that effective feedback requires that the LLSs have quite steep density profiles (see, e.g., the Appendix to \citealt{furl05-rec}).

More importantly, there is no obvious reason that the mean free path can \emph{only} change during overlap:  so long as (i) the emissivity continues to increase and (ii) $\VEV{\Gamma}$ controls the properties of LLSs, the feedback loop will continue -- and with the rapid increase in the collapsed fraction at $z \ga 6$, such a scenario seems entirely plausible.  Thus, even if $\VEV{\Gamma}$ is evolving strongly at $z \sim 6$ \citep{fan02, fan06}, we do not have a robust indicator of overlap.  

Indeed, it may be better to regard the final stage of reionization as the disappearance of LLSs \emph{after} overlap.  This ``post-overlap" phase, which consumes just a few percent of the neutral gas, is typically thought to be indistinguishable from the cosmic-web dominated Universe at later times.  We have shown instead that $\VEV{\Gamma}$ may evolve slowly during overlap but rapidly afterwards, making quasar absorption spectra a useful probe of this tail end of reionization (which matches smoothly onto the post-reionization Universe) but not necessarily of overlap itself.  This contrasts with the conventional picture in which $\VEV{\Gamma}$ evolves rapidly \emph{only} during overlap; that intuition came from reionization simulations that were too small to include the full inhomogeneity of reionization and so exaggerated the importance of overlap (e.g., \citealt{gnedin00}).  The recent semi-numeric calculations of \citet{choudhury08}, which included an approximate prescription for self-shielding gas, present a similar picture to ours, in which these LLSs are burned off over the (rather extended) final stages of reionization that follow overlap.

Beyond an increase in $\Gamma$, there are other reasons that $r_0$ may evolve throughout (and after) reionization.  For example, photoheating can evaporate loosely-bound structures, substantially modifying the IGM gas distribution by eliminating dense systems \citep{haiman01, shapiro04, pawlik08}.  This will increase $r_0$ as well; however, the process should happen gradually throughout reionization as more and more of the volume is illuminated (and so will be particularly slow if reionization is extended).  Evaporation also occurs over the sound-crossing time, which is relatively long for the moderate overdensities of most interest at these high redshifts \citep{pawlik08}, so that LLSs probably continue to evolve past the ``end" of reionization.

Given the challenges we have raised to the conventional interpretation, it is worth asking two further questions about the data.  First, are there other effects that can cause a rapid increase in $\tau_{\rm eff}$ \emph{without} a significant change in $\Gamma$?  One possibility is the density distribution itself:  at least according to current models, it is extremely steep in the low-density tail that is responsible for \lya forest transmission, so it is possible that a small increase in the ionizing background renders a relatively large fraction of the IGM visible \citep{oh05}.  This must await more detailed calculations of the evolving density field at the end of reionization.

Second, how certain can we be that overlap has actually occurred by $z \sim 6$?  Is it possible that, although most of the IGM is highly ionized before that point, a small fraction far from ionizing sources could still be completely neutral \citep{lidz07}?  In the conventional picture, in which $\VEV{\Gamma}$ evolves rapidly at overlap, such a conclusion can be easily dismissed.  However, our results suggest that overlap itself may be buried inside the smoothly evolving transmission at $z \la 6$ -- or it may have occurred at much higher redshifts.

Finally, we expect our conclusions to hold with even more force during helium reionization:  the clumpy IGM and enhanced recombination rate during that era makes attenuation more important \citep{furl08-helium, mcquinn08, bolton08} and high-energy photons can more easily create a nearly-uniform ionizing background, so the transition from bubble-domination to web-domination will be even smoother.

In summary, with our present knowledge of high-$z$ LLSs there is no compelling reason to associate evolution in $r_0$ or \lya forest transmission exclusively with overlap.  If indeed the quasar data shows a more rapid evolution in $\tau_{\rm eff}$ at $z > 6$ \citep{fan06}, this may indicate that we are seeing a rapid increase in $r_0$ that followed or even preceded overlap (note, however, that this measurement is itself controversial; \citealt{songaila02, songaila04, becker07}).   More detailed studies of the coupling between the ionizing background and the dense, neutral blobs that trap ionizing photons are needed before the next step -- associating such a jump with overlap itself -- can be taken securely.

We thank A. Lidz and S.~P. Oh for comments that greatly improved the manuscript.  This research was partially supported by the NSF through grant AST-0607470 to SF.  Support for this work was also provided by NASA through Hubble Fellowship grant \#HF-01222.01 to AM, awarded by the Space Telescope Science Institute, which is operated by the Association of Universities for Research in Astronomy, Inc., for NASA, under contract NAS 5-26555.

\bibliographystyle{mn2e}
\bibliography{Ref_composite}

\end{document}